\documentclass[twocolumn,superscriptaddress,prb,showpacs,amsmath]{revtex4}
\usepackage{amsmath}
\usepackage{graphicx}
\usepackage{dcolumn}
\usepackage{bm}
\usepackage{color}

\begin{document}
\title{Edge states of Spin-1/2 Two-Leg Ladder with Four-Spin Ring Exchange}
\author{Mitsuhiro Arikawa}
\affiliation{Institute of Physics, University of Tsukuba 1-1-1 Tennodai, Tsukuba Ibaraki 305-8571, Japan}
\author{Shou Tanaya}
\affiliation{Institute of Physics, University of Tsukuba 1-1-1 Tennodai, Tsukuba Ibaraki 305-8571, Japan}
\author{Isao Maruyama}
\affiliation{Graduate School of Engineering Science, Osaka University, 1-3 Machikaneyama-cho,
Toyonaka, Osaka 560-8531, Japan}
\author{Yasuhiro Hatsugai\footnote{corresponding author}}
\affiliation{Institute of Physics, University of Tsukuba 1-1-1 Tennodai, Tsukuba Ibaraki 305-8571, Japan}
\affiliation{
Kavli Institute for Theoretical Physics, University of California, Santa Barbara, California 93106, USA}%

\date{\today}
\begin{abstract}
A topological insulator and its spin analog as a gapped spin liquid 
have characteristic  low energy excitations (edge states) within the gap when
the systems have boundaries. This is the bulk-edge correspondence, which implies
that the edge states themselves characterize the gapped bulk spin liquid.
Based on the general principle, we analyzed the 
 vector chirality and rung singlet phases of the spin-$\frac{1}{2}$ ladder 
with ring exchange
by using the edge states and the entanglement entropy. 
\end{abstract}
\pacs{73.43.Nq,75.10.Jm,75.40.Cx}
\maketitle

\section{Introduction}
Spontaneous symmetry breaking has been
fundamental to characterize  phases of matter.
An example is a magnetic phase where
a local order parameter is defined as a 
 quantum or thermal average of
local combination of spins. 
Its ordered phase is not invariant against the symmetry operation
which leaves the Hamiltonian invariant.
In spite of enormous success of this concept, 
it has been also realized that there still exist many important phases
that cannot be captured by the spontaneous symmetry breaking. 
Strong quantum fluctuation in low dimensions 
prevents such formation of the ordered state and
makes it possible to realize
states without any fundamental symmetry breaking.
Such phases form a novel class of matter 
as  quantum liquids and spin liquids.
Typical examples of the quantum liquids are many of integer and fractional 
quantum Hall states.
Spin analog of the quantum liquids is the spin liquid which includes 
the  Haldane integer spin chain \cite{haldane83,affleck87},  generic
 valence bond solid (VBS) states\cite{arovas88,katsura2007}, and some of the exactly solvable models that also
have these spin liquid ground states (Kitaev model\cite{kitaev2006}, tensor category \cite{levin2006} and so on).
 
As for an excitation of a quantum system, local characters of the 
order parameter and the 
symmetry breaking are fundamental. Generically speaking, 
one requires some mechanisms to have a gapless excitation.
Typical machineries are the existence of the Fermi surface (formed by fermionic
quasiparticles) and 
the Nambu-Goldston bosons associated with the broken continuous symmetry.
An example of the latter is the spin wave excitation associated
with the Neel order.
Generically one may construct a gapless  excitation by
spatial modulation of the local order parameter such as 
the Lieb-Shultz-Mattis type spin twists\cite{lieb1961}. 
In contrast, as for the quantum liquids and the spin liquids, 
it is natural to have a finite excitation gap
unless one assumes exotic spinon Fermi surfaces.

To characterize such quantum liquids and the spin liquids,
topological quantities such as the Berry phases and the Chern numbers
are quite useful\cite{hatsugai2004,hirano2007a}. 
They do not have any symmetry breaking nor local order parameter. 
Still, there are many kinds of interesting quantum states 
without characteristic low energy excitation. 
Such non-trivial quantum phases are the topological 
insulator and its spin analog.  
A generic concept to discuss such a phase is the 
topological order which was 
first considered for the quantum Hall states\cite{wen89}. 
As in the quantum Hall 
states\cite{laughlin81,halperin82,hatsugai93,hatsugai93a},
 non-trivial topological insulators do have 
characteristic edge states or impurity states. The bulk is 
gapped and insulating.
However the existence of boundaries or impurities
brings low energy excitations. 
The quantum Hall state of graphene  
also belongs to this topological insulator\cite{hatsugai2006a,arikawa2008,hatsugai2008rev}. 
This is the bulk-edge correspondence where
topologically non-trivial bulk guarantees the existence of
localized modes and such low energy localized excitations characterize
the gapped bulk insulator\cite{hatsugai93} conversely.

Not only in the electronic systems, but also, in a quantum spin system
such as the Haldane spin chain, is
the bulk-edge correspondence realized as the existence of 
the Kennedy triplet
for an open chain\cite{kennedy90}, 
which was confirmed experimentally as well\cite{hagiwara90}. 
As a novel theoretical tool, the entanglement entropy (E.E.)
for the gapped topological insulators 
has been quite successful to classify the VBS states of the
spin chains\cite{katsura2007,hirano2007},
 where this bulk-edge correspondence plays a fundamental role. 

The spin-$\frac{1}{2}$ two-leg ladder with four-spin ring exchange has rich phase structure due to the frustration\cite{lauchili2003}.
For this model, the static and dynamical properties have been intensively studied\cite{lauchili2003,honda93,Schmidt03,Hikihara03,Haga02,nishimoto08,
Hijii03,song2006,maruyama2008}. 
Although the ring exchange model is simple, due to the frustration, the ground state is quite involved, namely 
in the vector-chirality (VC) phase.  
In this paper we consider the VC phase and the rung-single (RS) phase of this model to
identify the ground state properties from the view point of the bulk-edge correspondence. 
\section{Model}
\begin{figure}
\begin{center}
\includegraphics[width=3.2in]{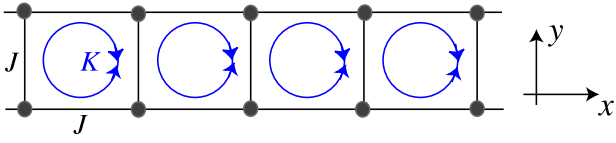}
\end{center}
\caption{Lattice structure: $J$ is the exchange interaction in the leg or rung direction and $K$ is the four-spin cyclic interaction. $x$-($y$-) axis is defined as the leg (rung) direction.\label{figlattice} }
\end{figure}

The Hamiltonian is given by
\begin{eqnarray}
{\mathcal H}_{\rm cyc} 
& = & J \left[ \sum_{x=1}^{N/2} \sum_{y=1,2} {\mathbf S}_{x,y} \cdot {\mathbf S}_{x+1,y}
+  \sum_{x=1}^{N/2}  {\mathbf S}_{x,1} \cdot {\mathbf S}_{x,2} \right] \nonumber \\
& & + K  \sum_{x=1}^{N/2} (P_x+P_x^{-1}).
\label{hamiltonian}
\end{eqnarray}
Hereafter, we parametrize the  exchange parameter $(J,K)$ as $(J,K)=(\cos \theta, \sin \theta)$.
The four spin cyclic exchange consists of the two and four spin exchange interactions as\cite{honda93},
\begin{eqnarray}
P_x+P_{x}^{-1} & = & {\mathbf S}_{x,1} \cdot  {\mathbf S}_{x,2}
+  {\mathbf S}_{x+1,1} \cdot  {\mathbf S}_{x+1,2} +  {\mathbf S}_{x,1} \cdot  {\mathbf S}_{x+1,1} \nonumber \\
& &
 +{\mathbf S}_{x,2} \cdot  {\mathbf S}_{x+1,2}
 +  {\mathbf S}_{x,1} \cdot  {\mathbf S}_{x+1,2} +  {\mathbf S}_{x,2} \cdot  {\mathbf S}_{x+1,1}  \nonumber \\
  & &+ 4 ({\mathbf S}_{x,1} \cdot  {\mathbf S}_{x+1,1} ) ({\mathbf S}_{x+1,1} \cdot  {\mathbf S}_{x+1,2} ) \nonumber \\
 & &+ 4 ({\mathbf S}_{x,1} \cdot  {\mathbf S}_{x+1,1} ) ({\mathbf S}_{x,2} \cdot  {\mathbf S}_{x+1,2} ) \nonumber \\
  & &- 4 ({\mathbf S}_{x,1} \cdot  {\mathbf S}_{x+1,2} ) ({\mathbf S}_{x,2} \cdot  {\mathbf S}_{x+1,1} ) + 1/4.
\label{cyclicexchange} 
\end{eqnarray}
Here, ${\mathbf S}_{x,y}$ denotes a spin-$\frac{1}{2}$ operator at site $(x,y)$ (see Fig.~\ref{figlattice}).
We assume the system-size $N$ to be even.

\begin{figure}
\begin{center}
\includegraphics[width=2.6in]{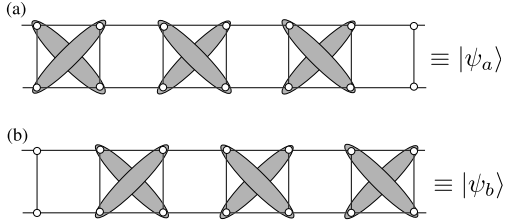}
\end{center}
\caption{Two types of the dimerized plaquette-singlets states. Shaded links denote the singlet of the two $S=1/2$ spins. \label{dimerstate}}
\end{figure}

\section{Edge states}
The Berry phases as topological order parameters classify the rung-singlet and the VC and RS phases of the model\cite{maruyama2008}.

In the VC phase, Hamiltonian (\ref{hamiltonian}) is adiabatically connected  to 
a decoupled vector-chiral model ${\mathcal H}_{\rm dvc} = \sum_{x= {\rm even}} ({\mathbf S}_{x,1} \times {\mathbf S}_{x,2}) \cdot 
({\mathbf S}_{x+1,1} \times {\mathbf S}_{x+1,2})$, whose ground state is a direct product of plaquette-singlets $| \psi_a \rangle $ [see Fig.~\ref{dimerstate}(a)], which is defined as
\begin{eqnarray}
| \psi_a \rangle & = &  [(1,1), (2,2)] \otimes  [(1,2), (2,1)] \nonumber \\
 & & \otimes [(3,1), (4,2)]  \otimes  [(3,2), (4,1)] \otimes \cdots
\end{eqnarray}
Here we have introduced a  singlet at two sites $\vec{\alpha}$ and $\vec{\beta}$, $[\vec{\alpha},\vec{\beta}]=[(\alpha_x,\alpha_y), (\beta_x,\beta_y)]$ as
\begin{eqnarray}
[\vec{\alpha},\vec{\beta}] & = & \frac{1}{\sqrt{2}} (|\uparrow \rangle_{\vec{\alpha}}|  \downarrow \rangle_{\vec{\beta}}-
| \uparrow \rangle_{\vec{\beta}} | \downarrow \rangle_{\vec{\alpha}} )
\end{eqnarray}
By taking into account of the translational invariance, we consider the
linear combination of
 $| \psi_a \rangle $ and 
 $| \psi_b \rangle $  as shown in Fig.~\ref{dimerstate}(b). State $| \psi_b \rangle$ is defined as
 \begin{eqnarray}
| \psi_b \rangle & = &  [(2,1), (3,2)] \otimes  [(2,2), (3,1)] \nonumber \\
 & & \otimes [(4,1), (5,2)]  \otimes  [(4,2), (5,1)] \otimes \cdots
\end{eqnarray}
 Notice that the two dimerized plaquette singlet states $| \psi_a \rangle $ and $| \psi_b \rangle $ are not orthogonal to each other, but their 
 overlap is exponentially small for large $N$.
For $N/2$ being even, the overlap is obtained as $\langle \psi_a | \psi_b \rangle = 2^{2-N/2}$.
As described later, the state $| \psi_s \rangle \propto | \psi_a \rangle + | \psi_b \rangle$ can be a good trial state to understand the edge states and the entanglement entropy
for the vector chirality state 
 \footnote{The phase factor can be arbitrary on $|\psi_b \rangle$.}.
	   
\begin{figure}
\begin{center}
\includegraphics[width=2.7in]{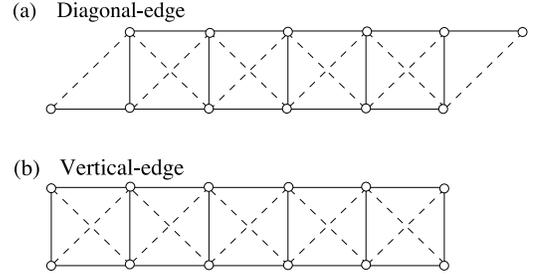}
\end{center}
\caption{Two types of open boundary conditions (OBCs).\label{howtocut}}
\end{figure}

\begin{figure}
\begin{center}
\includegraphics[width=2.4in]{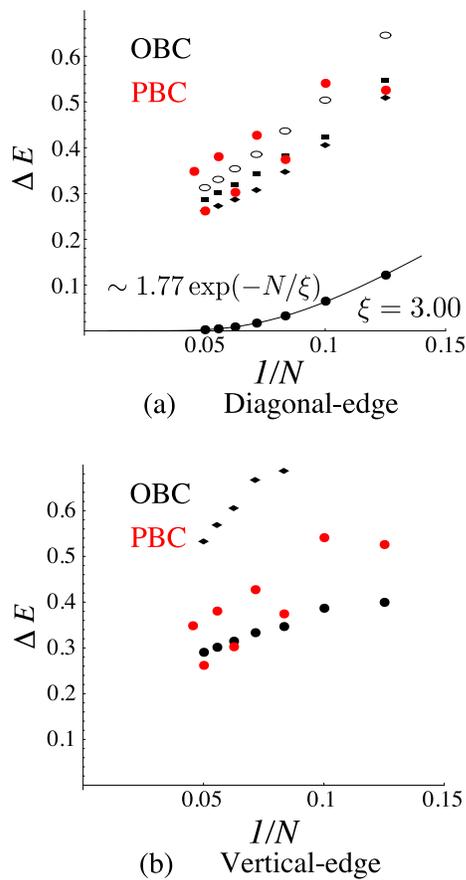}
\end{center}
\caption{System-size dependence of low excitation energy for $\theta=4\pi/5$ with the two OBCs
compared with the energy gap under the periodic boundary condition (PBC).
 \label{openenergygap}}
\end{figure}

\begin{figure}
\begin{center}
\includegraphics[width=2.8in]{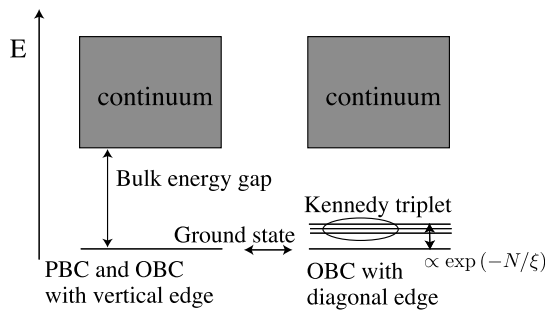}
\end{center}
\caption{Schematic picture of the low energy excitations.\label{schematic}}
\end{figure}

\begin{figure}
\begin{center}
\includegraphics[width=3.3in]{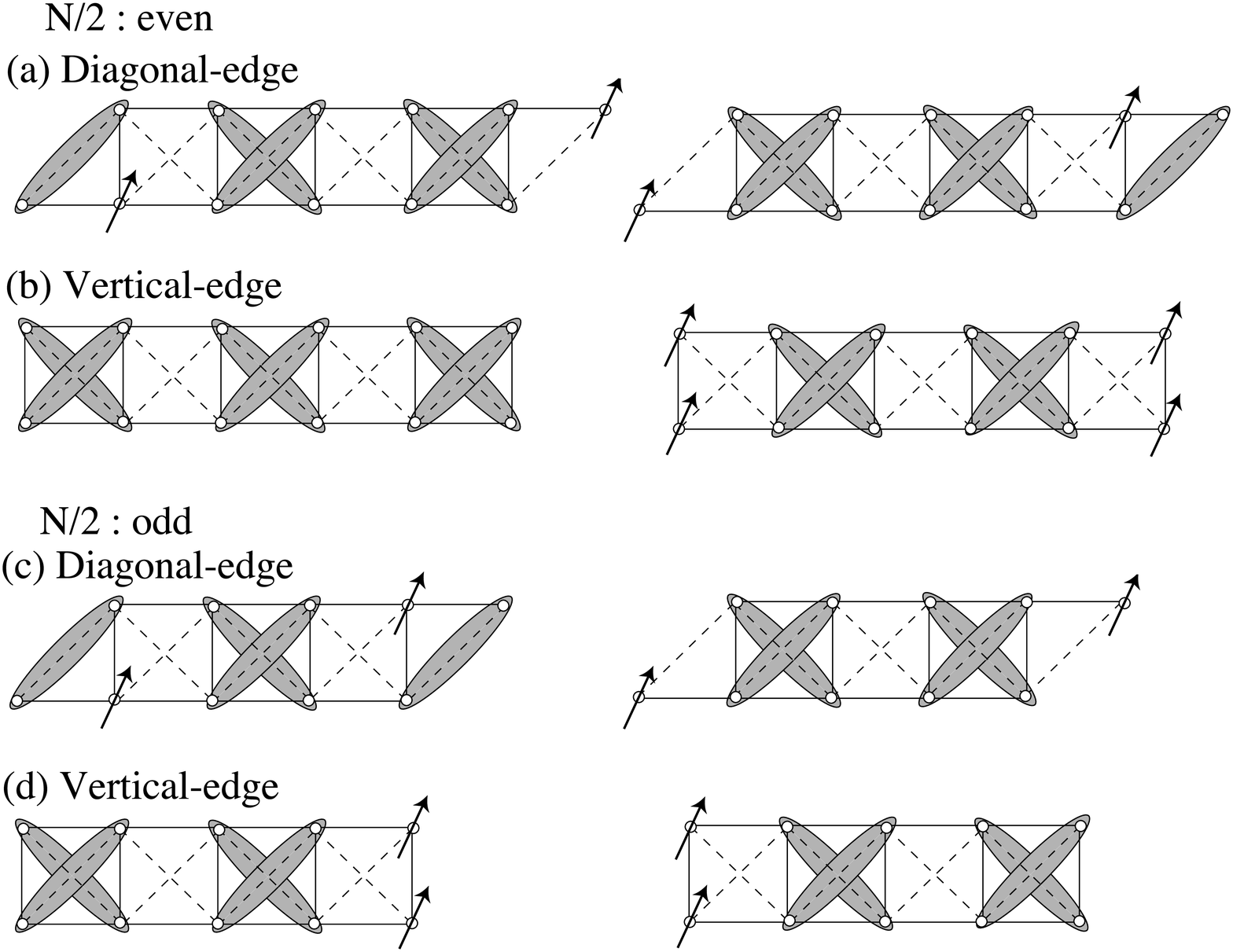}
\end{center}
\caption{Linear combination of the dimerized plaquette-singlets $| \psi_s \rangle$ with OBCs.\label{trial}}
\end{figure}

In the VC phase, the bulk itself has a finite gap in the thermodynamic limit\cite{lauchili2003,Hikihara03}. 
We introduce two types of boundaries ---(a) diagonal-edge and (b) vertical-edge (see Fig.~\ref{howtocut})
\footnote{We make a boundary by truncating all terms ${\mathbf S}_i \cdot {\mathbf S}_j$ and 
 $({\mathbf S}_i \cdot {\mathbf S}_j) ({\mathbf S}_k \cdot {\mathbf S}_l)$ whose bond
 cross the boundary. Note that some terms of 
 $P_x + P_x^{-1}$ in Eq.~(\ref{cyclicexchange}) do not cross the boundary even in the vertical-edge case.}.
 We diagonalized the Hamiltonian numerical by the Lanczos method. In Fig.~\ref{openenergygap} we show
 the size dependence on the energy gap of the total $S^z=0 $ sector for $\theta=4\pi/5$, 
 where the bulk spin gap is relatively large.
The system with the vertical-edges has almost the same excitation energy 
as that of the periodic one. In contrast, there exist additional 
low energy excited states with an exponentially small energy gap 
when the system has the diagonal-edges.
The appearance of such a  localized mode indicates 
the feature of the {\itshape bulk-edge correspondence}.
In fact, this mode is a triplet excitation and can be interpreted as the Kennedy triplet (see Fig.~\ref{schematic}) --- triplet excitation between the effective boundary spins at both sides \cite{kennedy90}. 
We confirmed that this mode is really a triplet excitation from the diagonalization on the different $S^z$ sector.
This mode is observed also in the different ($J,K$)'s in VC phase\footnote{Close to the phase boundary, the correlation length becomes large and the bulk spin gap is very tiny. 
In this case, "Kennedy triplet" is hardly observed.}.
Note that the Kennedy triplet in this model was also discussed in a different context before\cite{Hijii03}.  

When we introduce the boundary for the trial state $| \psi_s \rangle$, i.e., the 
linear combination of the two dimerized plaquette singlets, the isolated spins appear near the boundaries (see Fig.~\ref{trial}). 
In the system with diagonal-edges they appear at each boundary, 
while in the system with vertical-edges they appear as a pair at one side of the edge.
Although a pair of localized spins behaves freely in the decoupled vector chiral model ${\mathcal H}_{\rm dcv}$, the pair of the spins couples each other
 in the original model ${\mathcal H}_{\rm cyc}$ with vertical-edges. 
Therefore, the Kennedy triplet excitation does not appear in the system with the vertical-edges. 
Thus  the trial state $| \psi_s \rangle$ gives consistent understanding for the low energy spectra of the VC phase.

\begin{figure}
\begin{center}
\includegraphics[width=2.3in]{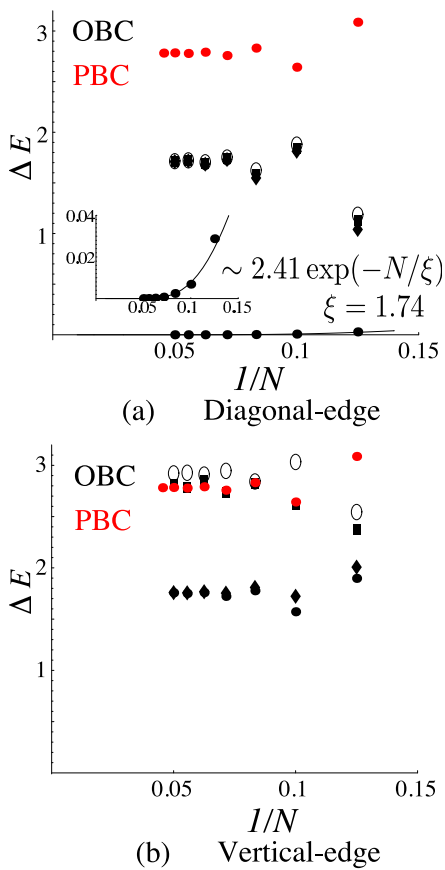}
\end{center}
\caption{System-size dependence of low excitation energy for $\theta=-\pi/5$ with the two OBCs
compared with the energy gap under the periodic boundary condition (PBC). The inset is the extended 
figure for the first excitation state energy. 
 \label{openenergygap2}}
\end{figure}

\begin{figure}
\begin{center}
\includegraphics[width=2.in]{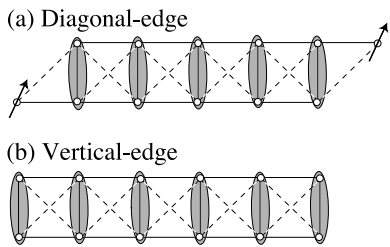}
\end{center}
\caption{Rung-singlets with OBCs.\label{trialR}}
\end{figure}

Next we consider the RS phase. In Fig.~\ref{openenergygap2}, we show the size dependence on the energy gap of the 
total $S^z=0$ sector for $\theta=-\pi/5$. Combining with the calculation on different $S^z$
sector we obtain the Kennedy triplet mode as in the VC phase in the case with diagonal-edge.
This is consistent with the naive picture of the rung singlet, which is obtained by the Berry phase (see Fig.~\ref{trialR}). This mode is confirmed also in the 
different ($J,K$)'s in the RS phase. 	

\section{Entanglement Entropy} 
\begin{figure}
\begin{center}
\includegraphics[width=3.2in]{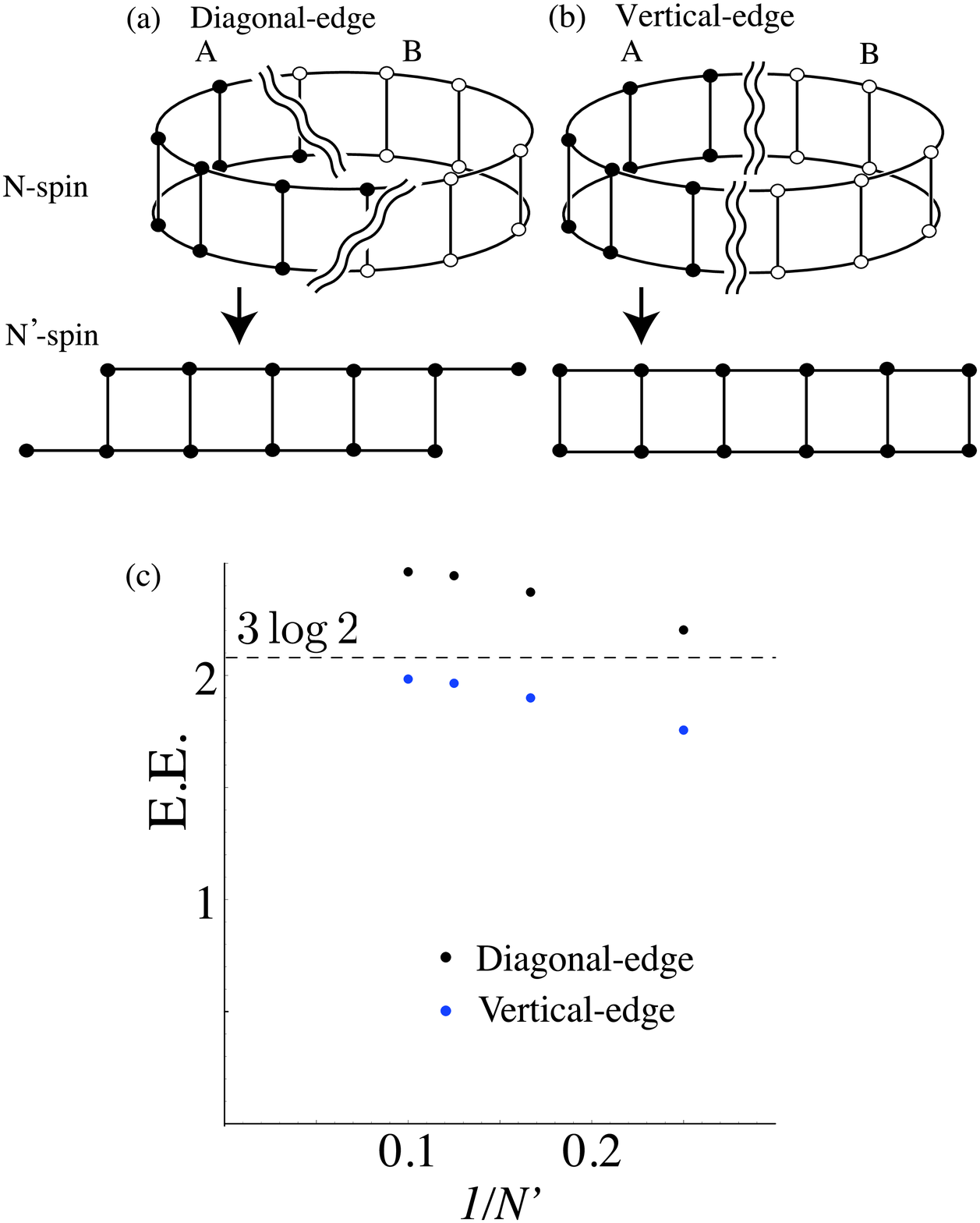}
\end{center}
\caption{Two types of the reduction to the subsystem $A$ with
(a) diagonal-edge and (b) vertical-edge. (c) Entanglement entropy of subsystem 
for $\theta=4\pi/5$ with $N'$ spins by reduction from $N=20$ spin system.
 \label{entanglement}}
\end{figure}

 In this section, let us consider entanglement entropy for the VC state.
 Define the density matrix of the state $|\psi \rangle$ as $\hat{\rho}=| \psi \rangle \langle \psi |$ and
divide the system into two subsystems $A$ and $B$ [see Figs.~\ref{entanglement}(a) and \ref{entanglement}(b)].
Then the entanglement entropy is defined as\cite{vidal2003} 
$$
{\rm E.E.}   =  -\langle \log \hat{\rho}_A \rangle_A = - {\rm Tr} \left[ \hat{\rho}_A \log \hat{\rho}_A\right]. 
$$
Here the reduced density matrix $\hat{\rho}_A$ is given as
 $\hat{\rho}_A=  {\rm Tr}_B \hat{\rho}$.
 It represents how much state $|\psi \rangle$ is entangled between subsystems
 $A$ and $B$. 
 
Let us assume state $|\psi \rangle$ as a unique ground state under the PBC 
 in the VC state. We take subsystem $A$ as the subsystem with the
diagonal (vertical) edges with $N'$ spins by the reduction from 
$N=20$ spin system with the PBC. Similarly as for the energy gap we consider two types of open boundaries, 
we calculate the entanglement entropy numerically for the cases with vertical-edge and diagonal-edge as shown in Figs.~\ref{entanglement}(a) and \ref{entanglement}(b). 
Figure \ref{entanglement}(c) shows the $N'$ dependence of the entanglement entropy. 
In both cases, the obtained E.E. contains a contribution around $3 \log 2$.
The  contribution $3 \log 2$ can be understood by the trial state $| \psi_s \rangle$ (see the Appendix).

In the naive picture, in the RS phase the isolated single on the rung is a good trial state. 
Therefore  we expect that E.E.$ = 2 \log 2$ for the diagonal edge and that E.E. $= 0$ for the vertical edge.
We have qualitative agreement that the E.E. for the  diagonal edge is larger than that for the vertical edge, although 
there is strong $\theta$-dependence and large deviation from the $2 \log 2$. 
This is due to the relatively large overlap on the leg-direction.

\section{Summary}
In summary, it has been shown that in the vector chirality
state the dimerized plaquette singlet state $|\psi_s \rangle$ can
be a good trial state
to
understand the numerical results for the topological properties --- 
the edge states
and the entanglement entropy. 
The entanglement entropy of $| \psi_s \rangle$ has been obtained as $3\log 2$
for both types of the reduced systems
while the appearance of the edge states depends on the type of boundaries.
These {\it boundary-dependent} low energy excitations
 as the generic edge states characterize the vector chirality phase.
This boundary dependent edge state is also observed in the rung-singlet phase. 
Such localized modes in the boundary are expected to be observed experimentally through 
the impurity or surface effect.

The work was supported in part by Grants-in-Aid for Scientific 
Research, Grant No.20654034 from JSPS and No.220029004 (Physics 
of New Quantum Phases in Super-clean Materials) and 
No.20046002 (Novel States of Matter Induced by Frustration) 
on Priority Areas from MEXT (Japan).
The work of YH was also supported in part by the National 
Science Foundation under Grant No. PHY05-51164.
Some numerical calculations were carried out on Altix3700BX2 at YITP in Kyoto University and 
the facilities of the Supercomputer Center, Institute for Solid State Physics, University of Tokyo.

\begin{figure}
\begin{center}
\vspace{0.5cm}
\includegraphics[width=3.2in]{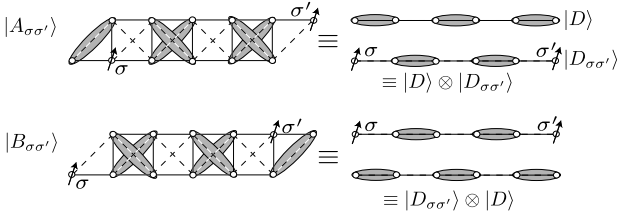}
\end{center}
\caption{Two types of the tensor products of the two dimerized states in Eq.~(\ref{example}) with diagonal-edges
for even $N'/2$ case ($N'=12$), which corresponds to Fig.\ref{trial}~(a).
 \label{entcalc}}
\end{figure}

\appendix
\section{Entanglement Entropy for the trial state }
For example, we calculate the entanglement entropy of $| \psi_s \rangle$
in the diagonal-edge case for even $N/2$ and $N'/2$ cases.
The reduced density matrix $\hat{\rho}_A$ can be obtained as (see Fig.~\ref{entcalc}),
\begin{eqnarray}
\hat{\rho}_A & = & \frac{1}{4m}\sum_{\sigma,\sigma'=\uparrow,\downarrow}
\left[ | A_{\sigma \sigma'} \rangle \langle  A_{\sigma \sigma'} | +
 | B_{\sigma \sigma' }\rangle \langle  B_{\sigma \sigma'}  | \right] \nonumber \\
  &   &+  \frac{m'}{4m}
  \left[ | A_{\uparrow \downarrow} \rangle \langle  B_{\uparrow \downarrow} |
 - | A_{\uparrow \downarrow}\rangle \langle  B_{\downarrow \uparrow}|
 + (A \leftrightarrow B)  \right],
\end{eqnarray}
where $m=2+2^{3-N/2}$ and $m'=2^{1-(N-N')/2}$. States $| A_{\sigma \sigma'} \rangle$ and $| B_{\sigma \sigma'} \rangle$
are represented as the tensor product of the dimerized states in the zigzag chains  (see Fig.~\ref{entcalc}),
\begin{eqnarray}
| A_{\sigma \sigma'} \rangle & = & 
[(1,1),(2,2)] \otimes | \sigma \rangle_{(2,1)} \nonumber \\
& &   \otimes [(3,1),(4,2)] \otimes  [(3,2),(4,1)] \otimes \cdots \nonumber \\
& &  \otimes [(N'/2-1,1),(N'/2,2)] \nonumber \\
& & \otimes  [(N'/2-1,2),(N'/2,1)] \otimes  | \sigma' \rangle_{(N'/2+1,2)}, \nonumber \\
& & \\
| B_{\sigma \sigma'} \rangle & = & | \sigma \rangle_{(2,1)} \otimes 
[(2,1),(3,2)]\otimes [(2,2),(3,1)] \otimes \cdots \nonumber \\
& &  \otimes [(N'/2-2,1),(N'/2-1,2)] \nonumber \\
& & \otimes  [(N'/2-2,2),(N'/2-1,1)] \otimes  | \sigma' \rangle_{(N'/2,2)} \nonumber \\
& & \otimes  [(N'/2,1),(N'/2+1,2)].
\end{eqnarray}
We introduce the dimerized states $| D \rangle$ and  $| D_{\sigma \sigma'} \rangle$ on the chain
with length $N'/2$,
\begin{eqnarray}
| D \rangle & = & [1,2] \otimes [3,4] \otimes \cdots \otimes  [N'/2-1,N'/2], \\
| D \rangle_{\sigma \sigma'} & = & | \sigma \rangle_1 \otimes  [2,3] \otimes \cdots \nonumber \\
& & \otimes  [N'/2-2,N'/2-1] \otimes 
 | \sigma' \rangle_{N'/2}.
 \end{eqnarray}
Changing site indexes to decouple the ladder into two chains, we have
equivalences
$| A_{\sigma \sigma'} \rangle = | D \rangle \otimes  | D \rangle_{\sigma \sigma'}$
and $| B_{\sigma \sigma'} \rangle =  | D \rangle_{\sigma \sigma'} \otimes | D \rangle $ (see Fig.~\ref{entcalc}). 
Here we have introduced the normalized state $| \tilde{D} \rangle \propto
 | D \rangle - \langle  D_{\uparrow \downarrow} | D  \rangle  | D_{\uparrow \downarrow} \rangle  - \langle  D_{\downarrow \uparrow } | D  \rangle  | D_{ \downarrow \uparrow} \rangle$ by the Gram-Schmidt orthogonalization method. Then we have the following relation:
 \begin{eqnarray}
\hat{\rho}_A   & \simeq  & \frac{1}{8} \sum_{\sigma,\sigma'=\uparrow \downarrow} \left[
\left(  | \tilde{D} \rangle  \otimes | D_{\sigma \sigma'} \rangle \right)
 \left( \langle \tilde{D} |  \otimes \langle D_{\sigma \sigma'} |  \right) \right. \nonumber \\
 & & \left.
 +
\left(   | D_{\sigma \sigma'} \rangle \otimes | \tilde{D} \rangle \right)
\left(   \langle D_{\sigma \sigma'} | \otimes \langle \tilde{D} |  \right)
 \right], 
 \label{example}
 \end{eqnarray}
The approximation holds up to the order ${\mathcal O}(2^{-N'/2}) + {\mathcal O}(2^{-(N-N')/2})$ and the eight summands correspond to the states
 shown in Fig.~\ref{trial}(a). 
 Thus, the entanglement entropy is obtained as ${\rm E.E.} \simeq - 8 \times \frac{1}{8} \log \frac{1}{8} = 3 \log 2$.
In a similar manner we can calculate the entanglement entropy in the case of the other geometry.  
The entanglement entropy counts the degrees of freedom of the spins around the edges,
although the Kennedy triplet does not appear in the system with vertical-edge due to the short range residual interaction between 
the localized effective spins.

\end{document}